\documentclass[%
 aip,
 amsmath,amssymb,
 reprint,%
]{revtex4-1}

\usepackage{amsmath}

\usepackage{graphicx}
\usepackage{dcolumn}
\usepackage{bm}
\usepackage{lineno}
\usepackage{lipsum}

\usepackage[utf8]{inputenc}
\usepackage[T1]{fontenc}
\usepackage{mathptmx}
\usepackage{etoolbox}
\usepackage{xcolor}
\usepackage{amsmath}
\usepackage{numprint}
\usepackage{bm}
\usepackage[T1]{fontenc}
\usepackage[hidelinks]{hyperref}
\usepackage{orcidlink}

\DeclareTextFontCommand{\edit}{\selectfont\color{red}}

\newcommand{\desgc}{DESGC}
\newcommand{\dspoc}{DSpOC}
\newcommand{\ud}{UD}
\newcommand{\udip}{UDIP}
\newcommand{\dapper}{DAPPEr}
\newcommand{\fdesgc}{Delaware Space Grant Consortium}
\newcommand{\fdspoc}{Delaware Space Observation Center}
\newcommand{\fud}{University of Delaware}
\newcommand{\fudip}{University of Delaware Ionospheric Probe}
\newcommand{\fdapper}{Delaware Atmospheric Plasma Probe Experiment}
\newcommand{\udesgc}{https://www.delspace.org/}
\newcommand{\delaware}{Delaware}
\newcommand{\ranyah}{\cite{khan2025delaware}}

\makeatletter
\def\@email#1#2{%
 \endgroup
 \patchcmd{\titleblock@produce}
  {\frontmatter@RRAPformat}
  {\frontmatter@RRAPformat{\produce@RRAP{*#1\href{mailto:#2}{#2}}}\frontmatter@RRAPformat}
  {}{}
}%
\makeatother
\begin{document}

\preprint{AIP/123-QED}

\title{A Student Sounding Rocket Experiment to Probe Electron Density in the Lower Ionosphere}

\author{Joshua Goodwill \orcidlink{0000-0002-5354-1164}}
\affiliation{University of Delaware, Newark, DE 19716, USA}
\email{goodwill@udel.edu}
\author{Timothy Tomaszewski \orcidlink{0009-0008-0997-0579}}
\affiliation{University of Delaware, Newark, DE 19716, USA}
\author{Jarrod Bieber \orcidlink{0009-0003-2814-2537}}
\affiliation{University of Delaware, Newark, DE 19716, USA}
\author{Carson P.~Brown \orcidlink{0009-0001-8945-4898}}
\affiliation{University of Delaware, Newark, DE 19716, USA}
\author{Ashton Southwick \orcidlink{0009-0001-2333-7816}}
\affiliation{University of Delaware, Newark, DE 19716, USA}
\author{Samuel Auerbach}
\affiliation{University of Delaware, Newark, DE 19716, USA}
\author{Sreenidhi Banda \orcidlink{0009-0009-3141-3207}}
\affiliation{University of Delaware, Newark, DE 19716, USA}
\author{Alex Bruce}
\affiliation{University of Delaware, Newark, DE 19716, USA}
\author{Marisol Catalan Olais \orcidlink{0009-0006-2442-6235}}
\affiliation{University of Delaware, Newark, DE 19716, USA}
\author{Markos Duey}
\affiliation{University of Delaware, Newark, DE 19716, USA}
\author{Ranyah Khan}
\affiliation{University of Delaware, Newark, DE 19716, USA}
\author{Connor McCleery \orcidlink{0009-0009-7857-0357}}
\affiliation{University of Delaware, Newark, DE 19716, USA}
\author{Jonathan Rosado \orcidlink{0009-0008-0750-3328}}
\affiliation{University of Delaware, Newark, DE 19716, USA}
\author{Amanda Swenson}
\affiliation{University of Delaware, Newark, DE 19716, USA}
\author{Matthew Ward}
\affiliation{University of Delaware, Newark, DE 19716, USA}
\author{Jarod Dagney \orcidlink{0009-0009-9457-2862}}
\affiliation{University of Delaware, Newark, DE 19716, USA}
\author{David Sundkvist \orcidlink{0000-0003-2794-7926}}
\affiliation{Space Sciences Laboratory, University of California, Berkeley, CA 94720, USA}
\author{John W.~Bonnell \orcidlink{0000-0002-0675-7907}}
\affiliation{Space Sciences Laboratory, University of California, Berkeley, CA 94720, USA}
\author{William H.~Matthaeus \orcidlink{0000-0001-7224-6024}}
\affiliation{University of Delaware, Newark, DE 19716, USA}
\author{Bennett A.~Maruca \orcidlink{0000-0002-2229-5618}}
\affiliation{University of Delaware, Newark, DE 19716, USA}



\date{\today}

\begin{abstract}
Earth's ionosphere is primarily generated by the Sun's ionization of atoms in the upper atmosphere.  This plasma environment is highly dynamic, exhibits strong spatial and temporal, variations and significantly impacts power systems, navigation, and satellite and ground communications.  The \fudip{} (\udip{}) was developed as a hands-on learning experience for \fud{} students on the ionosphere itself and on the development of space technology.  Students (mostly undergraduates) have led all aspects of the experiment including mission planning, electronics development, mechanical development, software development, and scientific data analysis. The present study is intended as a guide to readers on how the development of space and near-space flight missions can be successfully incorporated into an undergraduate physics curriculum. The fourth version of the experiment (\udip{}-4) flew along with student experiments from across the USA on the 2024 RockSat-C sounding rocket, which was launched from NASA Wallops Flight Facility.  During the flight, \udip{}-4's Langmuir probe successfully measured the ionosphere's density profile -- electron density as a function of altitude -- up to an apogee of $114~{\rm km}$.  Though the measured density profile agrees well overall with the predictions of theoretical models, \udip{}-4 identified a thin layer of enhanced electron density at an altitude of approximately $110~{\rm km}$ that is consistent with a sporadic E~layer ($E_{\rm{S}}$).
\end{abstract}

\maketitle


\section{Introduction} \label{sec:intro}
Earth's ionosphere is characterized by a substantial fraction of the neutral gas being ionized and extends from approximately $80$ to $\numprint{1000}~{\rm km}$ in altitude.  The ionized gas of the ionosphere absorbs, reflects, and refracts radio waves, which directly impacts many modern technologies: radio communication, navigation (including GPS), and satellite operations.  The ionosphere is effectively a transition layer between the terrestrial and space environments -- both a part of the Earth's atmosphere and a region heavily influenced by processes and plasma originating at the Sun.  Studying the ionosphere is critical for understanding space weather events that can cause major disruption and damage to power grids, communications systems, and spacecraft \citep{tech:nrc2008,tech:abt2017}.  Conditions in the ionosphere vary widely across space and time in very complex ways and has been the object of extensive and ongoing study.\citep{kelley2009earth}.

The observation and study of the lower portion of the ionosphere presents a particular challenge.  The gas at even the base of the ionosphere is too diffuse for conventional aircraft to achieve lift but also too dense for spacecraft to maintain stable orbits (due to drag).  The lower ionosphere can be remotely observed, e.g., via radio waves -- a technique that led to the discovery of the ionosphere in the early 20th Century\citep{russell2016space}.  For in-situ observations, sounding rockets provide the best method for bringing instruments into the lower ionosphere to directly measure its particles and fields\citep{christe2016_Introduction}.  Sounding rockets (also known as suborbital rockets) are medium-sized rockets that launch into space and fall back to Earth without achieving orbit that have a long history of exploring Earth's near space environment.

RockSat-C\footnote{https://www.nasa.gov/wallops/stem/rocksat} is an educational sounding rocket program run annually by NASA Wallops Flight Facility.  Each year, teams of students (mostly undergraduate) from across the USA spend the academic year designing and building a wide variety of experiments to fly on a sounding rocket.  The program is designed to provide students with access to low-cost flight opportunities, develop their engineering skills, and promote interest in STEM careers\citep{koehler2011rockon}.

The \fud{} (\ud{}), in partnership with the \fdesgc{}\footnote{\udesgc{}}, has a long history of participating in RockSat-C.  Successive teams of \ud{} students have developed and refined the \fudip{} (\udip{}), which features a Langmuir probe for measuring electron density.  Langmuir probes, which are among the oldest and most robust plasma instruments, are metal electrodes that measure the electrical properties of a plasma 
\citep{mott-smith1926_Theory, allen1992_Probe}.  When a voltage is applied to a Langmuir probe, a response current flows from the probe to the surrounding plasma.  Measuring the current for a series of voltages produces an I-V~curve, which can be fit with a response function to extract plasma parameters such as the electron density and temperature \citep{chen2001_Langmuir}.

This article summarizes the design, operations, and results of the fourth version of the \udip{} experiment (\udip{}-4), which flew into space on the 2024 RockSat-C sounding rocket and successfully measured electron density as a function of altitude in Earth's lower ionosphere.  It is intended to guide readers on how development of space and near-space flight missions can be successfully incorporated into an undergraduate physics curriculum.  Section~\ref{sec:log} describes how the development of \udip{}-4 was funded and integrated into a collegiate physics course. Section~\ref{sec:instr} overviews the \udip{}-4 hardware (both mechanical and electrical) developed by the students.  Section~\ref{sec:flight} provides a timeline of the rocket flight.  Section~\ref{sec:results} describes the analysis of \udip{}-4 data and reports key scientific results.  Section~\ref{sec:outlook} discusses the program's benefits to the student participants and comments on the outlook for the \udip{} program going forward.

\section{Logistics, Pedagogy, and Funding} \label{sec:log}

The primary objective of RockSat-C is to provide college students with hands-on learning opportunities in space science and engineering.  It is structured around the academic year and is designed to be readily integrated into an undergraduate curriculum.  Many of the schools that participate in RockSat-C do so annually and have their students complete the program as part of their required senior-design course.  A typical student team consists of between four to eight members and works under the guidance of at least one faculty mentor.

The RockSat-C program begins anew each September when each team files an ``intent to fly form.''  Then, over the course of the fall semester, each team develops and tests their experiment's design.  As part of this process, the students prepare and deliver three virtual presentations to the RockSat-C leadership that mirror those used in NASA's scientific missions: conceptual design review, preliminary design review, and critical design review.  These review presentations help students maintain a steady work schedule and enable them to receive design feedback from the RockSat-C leadership.After the conceptual design review in December, the mission design is finalized, and the team receives notification from RockSat-C as to whether the experiment will be manifested on the rocket.  Additional reviews are scheduled through the spring semester and culminate in the visual verification check in early June.  In late June, some members of the team then travel with their experiment to NASA Wallops to integrate their experiment into the rocket and observe the launch.

At \ud{}, each new \udip{} team is formed in the spring semester before the program starts.  To ensure that the team has a balanced set of skills, students are recruited from a variety of physics and engineering majors using email, posters, and announcements in classes.  Since student interest typically exceeds team capacity, all students are required to apply and interview for the program.  The current year's \udip{} team is heavily involved in recruiting and interviewing the next year's team.  Once the new \udip{} team is recruited, its members hold occasional, virtual meetings during the summer to build group cohesion and give students the opportunity to discuss background reading and preliminary designs.  Each team member is required to enroll in a ``research for credit'' course at \ud{} -- three credits in the fall semester and three in the spring.  This course credit reflects the time commitment (roughly ten hours per week) made by each student and provides a mechanism for ensuring accountability.  Student performance is evaluated quarterly and discussed in one-on-one meetings with the instructor.

The \udip{}-4 team consisted of twelve undergraduate students, one graduate student, and two mentors (one faculty and one staff).  The students were divided among five subteams: science, software, electrical, mechanical, and management.
\begin{itemize}
    \item The science subteam defined the mission and instrument requirements (e.g., determined the appropriate measurement cadence and physical dimensions of the Langmuir probes).  Prior to flight, these students used existing measurements to generate synthetic data to aid other subteams in their design choices.  After flight, they oversaw the analysis of actual flight data. 
    \item The software subteam led the development of the mission’s two key pieces of software.  The flight software ran on the mission’s flight computer and executed and recorded all measurements and operations during flight.  The ground software ran on ground computers after flight and processed flight data. 
    \item The electrical subteam implemented all electrical and electronic components of the mission.  These included the science board (containing the electronics for the Langmuir probes), the flight computer, and the power system. 
    \item The mechanical team oversaw the physical design and implementation of the mission.  This included the Langmuir probes themselves, dimensions and placement of all components, and ensuring the entire experiment would survive the harsh conditions of flight (e.g., extreme vibrations, temperature, and drag forces). 
    \item The management subteam established the mission’s timeline and ensured that all of the subteams were remaining on schedule.  These students also oversaw safety and compliance (e.g., ensuring that the experiment would meet all of the requirements for flying on the rocket), tracked the experiment’s inventory, and led preparations for all review presentations with NASA Wallops.
\end{itemize}

The students selected an overall lead for the mission as well as a lead for each subteam.  Each student was required to attend at least two meetings each week: a subteam meeting and a full-team meeting.  Students scheduled work sessions and topic-specific meetings as needed.  Owing to the size of the team and the complexity of the mission, a strong emphasis was placed on organization and documentation. The two-semester commitment that students made to the project had them participate in \udip{}-4 from its conceptual design at the start of the fall semester through to its final fabrication, integration, and testing at the end of the spring semester.  The June trip to NASA Wallops (which included the integration and launch of the rocket) was optional but encouraged.  After the launch, some students chose to cycle off the project, but many chose to stay to help analyze the \udip{}-4 flight data and continue developing the \dapper{} CubeSat mission.

The RockSat-C program is closely tied to the National Space Grant College and Fellowship Program\footnote{https://spacegrant.org/} (commonly called ``Space Grant''), which was established by Congress in 1988 to promote space science education and research.  The program supports 52 consortia across the United States: one in each state, the District of Columbia, Puerto Rico, and Guam.  Most schools that participate in RockSat-C receive logistical, technical, and/or financial support to do so from their local Space Grant consortium.  In 2024, the program fee for a RockSat-C experiment (which included launch expenses) was \$\numprint{9000} for a half-canister load (such as \udip{}) and \$\numprint{15000} for a full-canister payload.  For \udip{}, Delaware Space Grant Consortium annually provides the workshop fee, up to \$\numprint{2000} for supplies, and travel and lodging expenses for the trip to NASA Wallops by the students and mentors.

\section{Instrumentation} \label{sec:instr}

\begin{figure}
\centering
\includegraphics[width=3in]{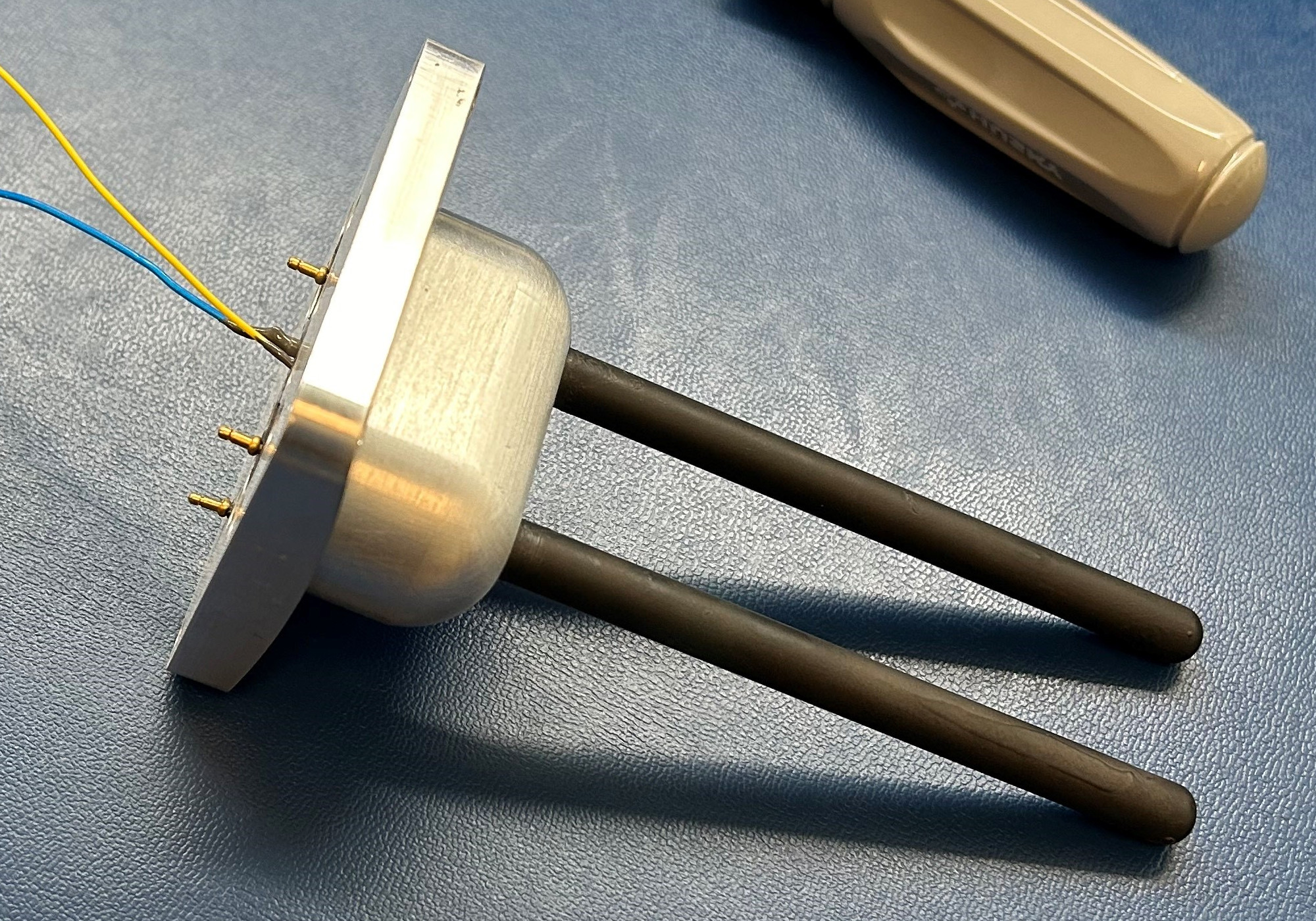} \\
\includegraphics[width=3in]{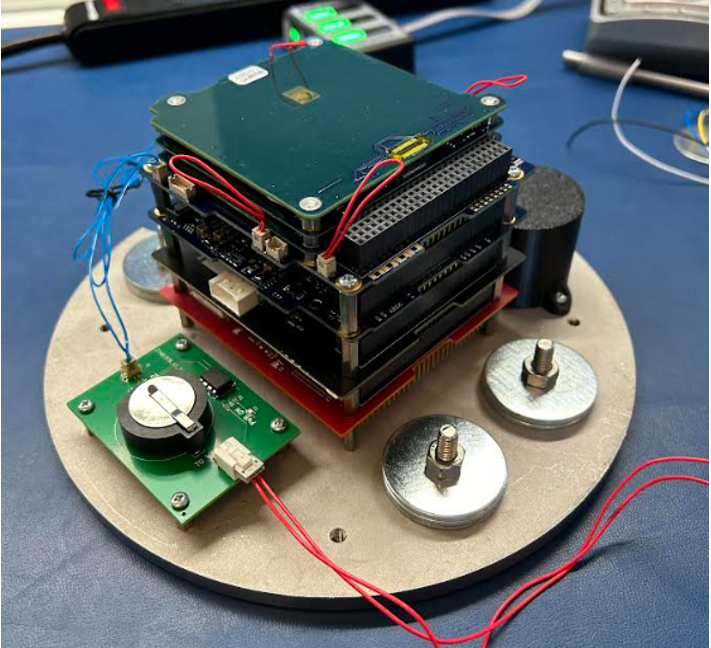}
\caption{\label{fig:instr} Photographs of the \udip{}-4 Langmuir probes (top) and electronics (bottom) while undergoing testing prior to integration onto the rocket.}
\end{figure}

Figure 1 summarizes the key mechanical and electrical systems of \udip{}-4..  The \udip{}-4 Langmuir probes (Section~\ref{sec:instr:probes}) were mounted to the exterior of the rocket while the stack of electronics boards (Sections~\ref{sec:instr:sci} to~\ref{sec:instr:power}) was mounted inside the rocket. Though this article offers an overview of the \udip{}-4 electronics, circuit diagrams and discussions of the theory of operation of the mission's custom electronics are provided elsewhere\ranyah{}.

\subsection{Langmuir Probes} \label{sec:instr:probes}

\udip{}-4 carried a pair of identical, cylindrical Langmuir probes each $3.75~{\rm in}$ long and $0.32~{\rm in}$ in diameter.  One probe served as the ``ground'' probe and was kept at $0~{\rm V}$ relative to the rocket's electrical ground; the other was the ``sweep'' probe and was swept through a series of DC voltages.  The current flowing from the sweep probe, through the surrounding plasma, and to the ground probe was measured for each voltage applied to the sweep probe (Section~\ref{sec:instr:sci}).  These measurements of current versus voltage constituted the I-V~curves that were \udip{}-4's fundamental data product, which are discussed further in Section~\ref{sec:results:exemplar}.

Special environmental considerations were included in the design of the \udip{}-4 Langmuir probes.  The 2024 \textit{RockSat-C User’s Guide} and program welcome documents provide some details on the mechanical environment encountered during flight.  The peak spin rate of the payload is typically $5.6~{\rm Hz}$.  Sustained accelerations of about $25g$ are typically encountered during ascent, and designing experiments to withstand impulses of up to $50g$ is recommended. For drag, the maximum value of $\rho v^2/2$ (where $\rho$ is the density of air around the rocket and $v$ is the rocket’s speed) is given as about $2.6 \times 10^5~{\rm Pa}$.  Per the RockSat-C User’s Guide, the full rocket payload (prior to flight) undergoes both sine-wave and random vibrational testing spanning frequencies from $10$ to $2000~{\rm Hz}$.

To endure the extreme drag forces of rocket flight, each probe was machined from H13 tool steel, which has a high yield strength.  Likewise, the electrical isolation of each probe from the rocket was maintained by a commercially fabricated short tube made from synthetic sapphire.  To prevent oxide formation (which would affect electrical performance), the exposed portion of each probe was coated with a colloidal graphite solution (AquaDAG). While Langmuir probes have been used on many different space missions (including sounding rockets), the specific designs vary considerably \citep{abe2013_Langmuir}.  Even just within sound-rocket missions, many different sizes, shapes, and electronics have been used for Langmuir probes.  Factors like the type of rocket, its specific trajectory, and where/how the probe is mounted on the rocket (e.g., a fixed probe versus a deployable one) all have major impacts on the design. For \udip{}-4, the students conducted a literature review to examine past Langmuir probe designs and heavily considered the RockSat-C user guide and past \udip{} missions to develop and effective design.

\subsection{Science Board} \label{sec:instr:sci}

The \udip{}-4 science board was a student-designed custom printed circuit board (PCB). This board hosted all of the analog and digital electronics required to generate I-V~curves from the Langmuir probes (Section~\ref{sec:instr:probes}).  A microcontroller on the science board used an external 12-bit digital-to-analog converter (DAC) to sweep through a series of voltages (ranging up to $-12$ to $+12~{\rm V}$) on the sweep probe.  A transimpedance amplifier was used to apply each sweep voltage to the sweep probe and then convert the resultant current to a voltage, which was measured by the microcontroller's analog-to-digital converters (ADC's).  The microcontroller carried out a voltage sweep on command from the flight computer (Section~\ref{sec:instr:fc}) and sent the resultant I-V~curve data to it via a serial communications line.

\subsection{Flight Computer} \label{sec:instr:fc}

The \udip{}-4 flight computer, also a student-designed custom PCB, was built around the industrial-rated Arduino Portenta H7 Lite microcontroller.  This board also included a nine degree of freedom (9-DoF) module for determining the payload's motion and orientation (via vector measurements of acceleration, angular velocity, and magnetic field) and an SD card module for recording all flight data.  The flight computer was programmed to collect and record data in three successive phases:
\begin{enumerate}
\item Ascent phase (before T+50~seconds).  The Langmuir probe system remained powered off, and high-cadence measurements were recorded from the 9-DoF module to characterize the rocket's launch.
\item Science phase (T+50 to +400~seconds).  Collection of each Langmuir probe I-V~curve was alternated with measurements from the 9-DoF module.
\item Descent phase (T+400 seconds onward).  Low-cadence measurements were recorded from the 9-DoF module, and the Langmuir probe system was powered off to prevent any potential electrical short between the probes when the rocket payload landed in the ocean.
\end{enumerate}

\subsection{Power System}\label{sec:instr:power}

Electrical power for all of \udip{}-4's systems was stored and distributed by the Clyde Space OPTIMUS~30 battery and STARBUCK-NANO electrical power system.  Though these components provided far more power storage and command options than needed for \udip{}-4, using them provided valuable experience (for students and mentors alike) for the development of future projects (Section~\ref{sec:outlook}).

\section{Rocket Flight}\label{sec:flight}

\begin{figure}
\centering
\includegraphics[width = \linewidth]{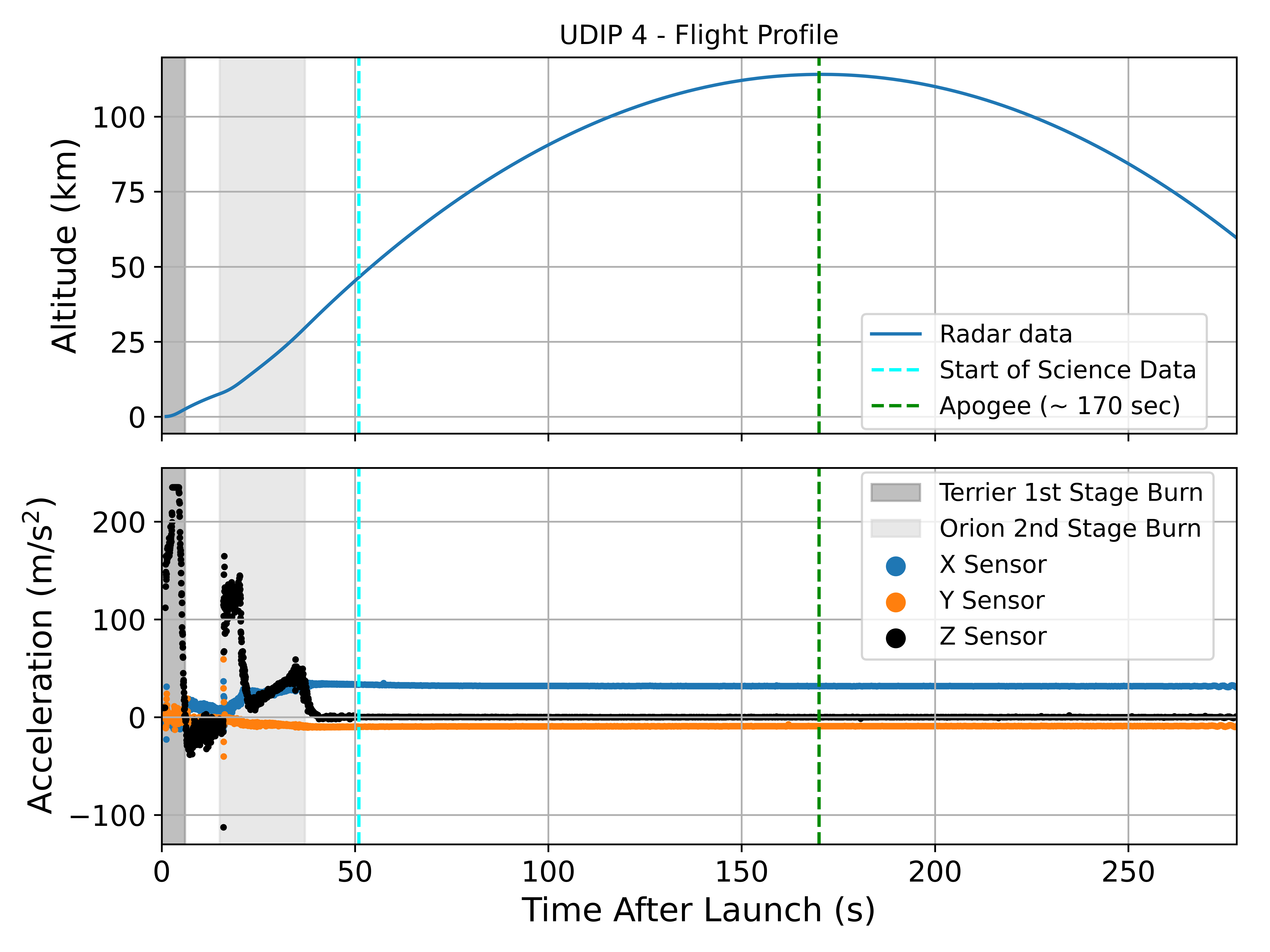}
\caption{\label{fig:Accel_Alt_vs_Time} Top: Altitude vs time after launch. Bottom: Acceleration vector vs time after launch. The darker shaded gray region indicates the Terrier $1\rm{st}$ stage burn, while the lighter shade is the Orion $2\rm{nd}$ stage burn. The blue vertical line denotes the beginning of \udip{}-4's Science data. The green vertical line indicates the time of apogee of the sounding rocket.}
\end{figure}

\udip{}-4 was launched on RockSat-C's Terrier-Improved Orion sounding rocket from NASA’s Wallops Flight Facility ($37^{\circ}\,50'\,21''\,{\rm N}$, $75^{\circ}\,28'\,54''\,{\rm W}$) on Thursday, June 20, 2024 at 05:30 EDT (09:30 UTC).  The rocket followed a nearly vertical trajectory and remained illuminated by dawn sunlight throughout its flight.  Figure~\ref{fig:Accel_Alt_vs_Time} shows the rocket's altitude and vector acceleration as functions of time after launch.  Altitude measurements were derived from radar tracking data from NASA Wallops Flight Facility.  Acceleration measurements (${\bf a} = a_x {\bf \hat{x}} +a_y {\bf \hat{y}} +a_z {\bf \hat{z}}$) were recorded from \udip{}-4's 9-DoF module (Section~\ref{sec:instr:fc}).  This module was mounted in the rocket's non-inertial reference frame with its $z$-axis oriented nose-ward along the rocket's axis.

At T+0 seconds, liftoff occurred with the ignition of the rocket's Terrier motor.  This first stage burned for approximately $5$~seconds, during which the peak upward acceleration was $a_z = 235~{\rm m}/{\rm s}^2 = 24 g$.  From T+5 to +15~seconds, the rocket coasted and slightly decelerated (i.e., $a_z<0$) due to drag.  At about T+15~seconds, the Improved Orion motor in the rocket's second stage ignited and then burned for about 25~seconds.  To provide the rocket with ``spin stabilization,'' the rocket was spun about its axis during this period, which is evidenced by an increase in centrifugal acceleration, $\sqrt{a_x^2+a_y^2}$.

After T+40 seconds, the rocket followed a ballistic trajectory.  It achieved apogee at approximately T+170~seconds at an altitude of $114~{\rm km}$.  The rocket payload deployed its parachute during its descent, splashed down into the Atlantic Ocean, recovered and returned to Wallops Flight Facility later the same day.

The \udip{}-4 electronics, which had been activated at approximately T-57 seconds, continued collecting data from the 9-DoF module until the recovered experiment was deactivated.  As noted in Section~\ref{sec:instr:fc}, the \udip{}-4 flight computer was programmed to only collect Langmuir probe data between T+50 and T+400~seconds, when the rocket was expected to be at its highest altitudes.

\section{Results} \label{sec:results}

This section summarizes the scientific results from the \udip{}-4 flight and includes samples flight data, a description of data processing, and the final flight results.


\subsection{I-V~Curve} \label{sec:results:exemplar}

\begin{figure}
\centering
\includegraphics[width=\linewidth]{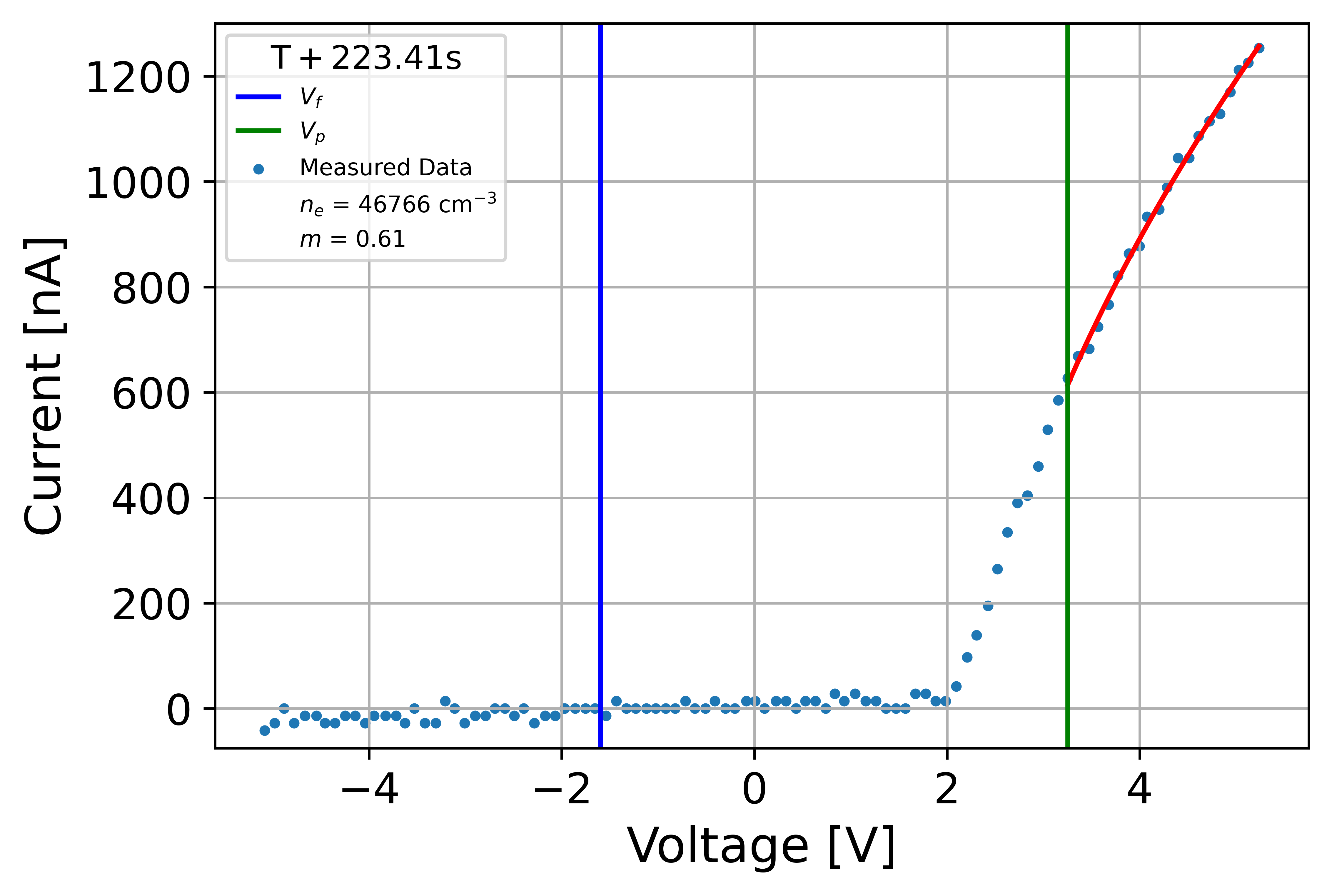}
\caption{\label{fig:exemplar} (Color online) I-V~curve (i.e., plot of current versus probe voltage) measured by \udip{}-4 at T+223.41 seconds.  Data are shown as cyan points, and vertical lines indicate the floating potential ($V_f$, blue) and plasma potential ($V_p$, green).  The best-fit curve of the data from the electron saturation ($V>V_p$) is plotted in red.  See Sections~\ref{sec:results:exemplar} and~\ref{sec:results:analysis} for full definitions and analysis.}
\end{figure}

Figure~\ref{fig:exemplar} shows an example I-V~curve collected by \udip{}-4 shortly after reaching apogee.  To generate this curve, the science board (Section~\ref{sec:instr:sci}) swept the probe voltage, $V$, from about $-6$ to about $+6~{\rm V}$ in equally spaced steps.  For each voltage step, the science board measured the current, $I$, that flowed through the plasma.  The measured $(I-V)$ data appear in Figure~\ref{fig:exemplar} as cyan dots.  The floating potential ($V_f$; the voltage for which the smoothed $I=0$) and the plasma potential ($V_p$; voltage at maximum $\frac{dI}{dV}$) are indicated with blue and green vertical lines, respectively.


The data with $V>V_p$ are in the ``electron saturation region'' and are characterized by currents that arise almost entirely from electron motion.  The response function in this region is\citep{chen2001_Langmuir,shunko2008_Langmuir,oyama2013introduction}:
\begin{equation} \label{eq:electron saturation}
    I = e n_e A \sqrt{\frac{k_{\rm B}T_e}{2 \pi m_e}}\left[ 1 + \frac{e(V-V_p)}{k_{\rm B}T_e}\right]^m ,
\end{equation}
where $m = 1/2$ for a long, thin cylindrical probe, but $m=1$ for a spherical probe.

\subsection{Fitting for Electon Density} \label{sec:results:analysis}

An estimate of the electron density, $n_e$, can be extracted from each I-V~curve for \udip{}-4 data using a non-linear fitting algorithm to fit Equation~\ref{eq:electron saturation} to the points from the electron saturation region ($V>V_p$).  Each I-V~curve's $V_p$-value was identified by smoothing the data and then determining the voltage for which the current gradient was highest.  Figure~\ref{fig:exemplar} shows the best-fit curve for the example I-V~curve in red.

In initial fits of the I-V~curves, $T_e$ was allowed to vary as a free parameter along with $n_e$, but the fitting algorithm consistently returned uncertainties in $T_e$ that were multiple orders of magnitude larger than the best-fit values in $T_e$.  The problem in reliably extracting electron temperature measurements from Langmuir probe data is a common one in the lower ionosphere\citep{hirao1972_Critical,rudakov2004_Effect}, where the gas's ionization fraction remains relatively low.  Consequently, a fixed value of $T_e=500~{\rm K}$ was chosen for all fits based on theoretical predictions \citep{bilitza2022_International} and measurements from other missions\citep{oyama2013introduction}.

Conversely, the exponent $m$ in Equation~\ref{eq:electron saturation} was allowed to vary as a free parameter in each I-V~curve fit.  The value of $m=1/2$ assumes that the probe is a long, thin cylinder, and the \udip{}-4 probes were made shorter and thicker than ideal to survive intense drag forces during launch (Section~\ref{sec:instr:probes}).  Even so, for most I-V~curves, the best-fit $m$-value was closer to $m=1/2$ than to $m=1$ (the value for a spherical probe). 

\subsection{Electron Density Profile} \label{sec:results:profile}

\begin{figure}[!htb]
\centering
\includegraphics[width=\linewidth]{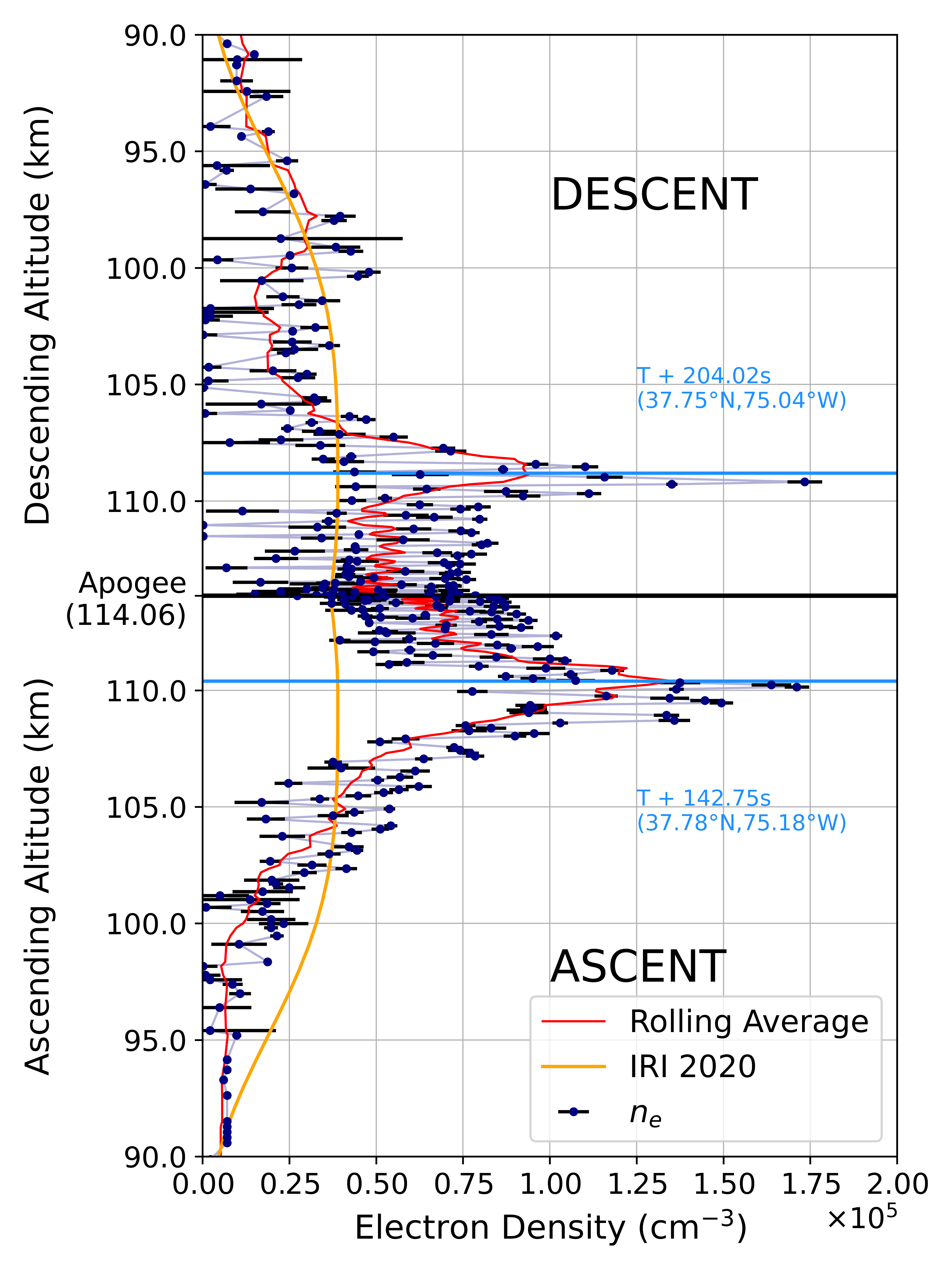}
\caption{\label{fig:Dens_vs_Altitude} (Color online) Electron number density, $n_e$, as a function of altitude for both the rocket's ascent (bottom) and descent (top).  Values for $n_e$ (including error bars) from the \udip{}-4 I-V~curves are shown in dark blue.  The red curve shows a ten-point rolling average of these data, and orange curve shows the predictions of the IRI model \citep{bilitza2022_International} for comparison.}
\end{figure}

Figure~\ref{fig:Dens_vs_Altitude} shows the trend in \udip{}-4 measurements of $n_e$ versus altitude for both the rocket's ascent and descent.  Data from altitudes less than $90~{\rm km}$ are omitted since the electron density was generally too low for \udip{}-4 to reliably measure.  A smoothed version of the \udip{}-4 data (produced with a ten-point smoothing window) is plotted in red. The error bars indicate the uncertainties in density returned by the fitting algorithm and do not include any additional error resulting from the choice of a fixed electron temperature (Section~\ref{sec:results:analysis}).

The un-smoothed data show a quasi-periodicity that mostly likely is the result of the rocket's spin.  As the rocket rotated, the Langmuir probe passed in and out of sunlight, which caused the number of photoelectrons around it to vary widely.  This interpretation is supported by a Lomb-Scargle periodogram \citep{vanderplas2018_Understanding} that was produced from the time series of $n_e$ measurements and that showed a strong peak at $5.4~{\rm Hz}$. This frequency corresponds well with the rocket's average measured spin rate.  Though the complex effects of sunlight on Langmuir probes certainly affect the absolute calibration of \udip{}-4 density measurements, the relative changes in $n_e$ values are likely robust.

The orange curve in Figure~\ref{fig:Dens_vs_Altitude} shows the theoretical electron density profile predicted by the International Reference Ionosphere (IRI~2020) model\citep{bilitza2022_International} for the date, time, and location of the rocket launch.  Overall, the IRI values agrees well with the observed values except for a relatively narrow layer of enhanced electron density that \udip{}-4 observed both on its ascent and descent at an altitude of roughly $110~{\rm km}$.  This feature is consistent with a sporadic E~layer (${\rm E}_{\rm S}$)\citep{haldoupis2012_Midlatitude,yu2019_global,oyama2008_Electron, mathews1998_Sporadic}.  \udip{}-4 observed the altitude and thickness of the layer to be slightly different on the ascent and descent.  The peak density occurred on the ascent at an altitude of about $111~{\rm km}$ but at $108~{\rm km}$ on the descent.  Likewise, the thickness (full-width at half-maximum) of the layer was about $4~{\rm km}$ on the ascent and $2.5~{\rm km}$ on the descent.  Most likely, these differences are mostly due more to spatial inhomogeneities than to temporal changes since \udip{}-4's two passages through the layer were separated by $13~{\rm km}$ in horizontal distance but only $62~{\rm s}$ in time.

The most widely accepted and supported theory for the $E_{\rm{S}}$ layer is the wind shear theory, resulting from vertical electric field induced motion occurring from shear layers in the ionospheric plasma \citep{whitehead1970_Production, whitehead1989_Recent, layzer1972_Theory}. The shear is developed through an enhancement in neutral density at roughly $125~\rm{km}$ altitude, which tends to carry metallic ions with the zonal neutral motion \citep{mathews1998_Sporadic} and thus forces ions into a thin dense sheet. In previous sounding-rocket missions launched from Wallops Island, these layers were found to occur between $100$ and $110 ~\rm{km}$ with a density enhancement between roughly $3$ and $6$ times the background measurements \citep{smith1972_Rocket}. Additionally, \udip{}-4 displayed a convergence in both electron density and altitude of the $E_{\rm{S}}$ layer with a rocket that was launched at a similar time in a series of night-time sounding rockets also launched from Wallops Island \citep{smith1970_sequence}. While \udip{}-4 did not have a secondary measurement device for confirmation of the $E_{\rm{S}}$ layer, the previous literature agrees well with our measurements. This indicates that the electron density was measured correctly, and that the deviations from the IRI are in fact a physical phenomenon in the ionosphere.

\section{Conclusions and Future Work}\label{sec:outlook}


Designing \udip{}-4 and analyzing its data provided students with real-life experience in all the essential areas of an undergraduate physics curriculum: mechanics (ballistics and drag), electromagnetism, thermodynamics, quantum mechanics (photoelectric effect), electronics, and data analysis.  The \udip{}-4 students also gained skills in collaboration, documentation, team organization, and time management, all of which are essential for any flight mission.  Already, some of the \udip{}-4 students have graduated and gone onto graduate schools and careers in academia and industry.

\udip{}-4 also served a critical role as the test flight of the \fdapper{} (\dapper{}).  This ambitious, student-led CubeSat mission is slated to deploy into orbit from the International Space Station in 2027 and become \delaware{}'s first orbital spacecraft.  Like \udip{}-4, \dapper{} will carry a pair of Langmuir probes, but it will use them to explore the upper ionosphere instead.  \dapper{} is drawing on substantial heritage from \udip{}-4 -- most notably its power system, flight computer, and science board.  Additionally, many \udip{}-4 students have continued to work on \dapper{}, bringing them essential technical and management experiences. 

\section{Acknowledgments}

The authors extend their deepest thanks to Christopher Koehler, Joyce L. Winterton, and everyone associated with the RockSat-C program and NASA Wallops Flight Facility for all their assistance in developing, integrating, launching, and recovering the \udip{}-4 experiment.  Special thanks also go to Angela Miller (\fdesgc{}) for logistical support and Derrick Allen (\fud{}) for machining services.  The authors gratefully acknowledge the help of Michael Wiltberger (University Corporation for Atmospheric Research; UCAR) in interpreting Figure~\ref{fig:Dens_vs_Altitude}.  Financial support for \udip{}-4 was provided by the \fdesgc{} (\desgc{}) under NASA Award~80NSSC20M0045.  Laboratory space and equipment were provided by the \fdspoc{} (\dspoc{}), which is supported by NASA Awards~80NSSC22K0884 and~80NSSC23K0608.

\section*{Author Contributions}

Data analysis and drafting of this article was lead by the mission's student science team, which included authors JG, JB, CPB, and AS.  Mentorship to the students was provided by authors JD, DS, JWB, WHM, and BAM.  All other authors were students who contributed to the mission's hardware and/or software development.

\bibliography{DAPPEr}
\end{document}